\documentclass[a4paper,11pt]{article}
\usepackage{pos}

\newcommand{\Eq}[1]{Eq.~\eqref{#1}}

\newcommand{\Refs}[1]{Refs.~\cite{#1}}

\newcommand{\Fig}[1]{Fig.~{\ref{#1}}}

\newcommand{\Table}[1]{Table~\ref{#1}}

\renewcommand{\thefootnote}{\arabic{footnote}}

\newcommand{\CDOT}{\mbox{$ \cdot $}}
\newcommand{\pslash}{\mbox{$\not \! p$}}
\newcommand{\aslash}{\mbox{$\not \! a$}}
\newcommand{\nslash}{\mbox{$\not \! n$}}

\newcommand{\Dslash}{\mbox{$\not \! \!D$}}

\newcommand{\Rpartial}{\overset{\rightarrow}{\partial}\!\!\!\!\phantom{\partial}}
\newcommand{\Lpartial}{\overset{\leftarrow}{\partial}\!\!\!\!\phantom{\partial}}
\newcommand{\LRD}{\overset{\leftrightarrow}{D}\!\!\!\!\!\phantom{D}}
\newcommand{\RD}{\overset{\rightarrow}{D}\!\!\!\!\!\phantom{D}}
\newcommand{\LD}{\overset{\leftarrow}{D}\!\!\!\!\!\phantom{D}}

\title{Gluon Parts of Gravitational Form Factors and Mass Distribution}

\author{Peter~C.~Tandy}
\affiliation{Center for Nuclear Research, Department of Physics, Kent State University, Kent OH 44242 USA}

\emailAdd{tandy@kent.edu}

\abstract{The parton structure of the nucleon and pion is investigated in an exploratory  model that allows one to assess whether the dressing of quarks can, by itself, produce realistic gluon contributions to light-cone momentum fractions, gravitational form factors, mass/energy distributions and their radii. The model is the Dyson-Schwinger Equations in Rainbow-Ladder truncation.  For the parton mass/energy distributions as a function of momentum transfer, we directly calculate  matrix elements of the Energy-Momentum Tensor by utilizing its similarity to the momentum fraction moment of GPDs associated with deep inelastic scattering. A variety of gravitational form factors are obtained including the D-term.}

\FullConference{The XVIth Quark Confinement and the Hadron Spectrum Conference (QCHSC24)\\
 19-24 August, 2024\\
 Cairns Convention Centre, Cairns, Queensland, Australia\\}

\begin{document}
\maketitle

\section{Introduction}
The gluon contributions to hadron properties and structure have long received much less attention than traditional quark contributions.  That situation has changed in recent years due to developments in  both experiment and theory.  When the attention to hadron structure broadened to include not only internal electromagnetic properties and parton distribution functions (PDFs), but also internal distributions of energy, angular momentum, and mechanical properties  then the hadron matrix elements of the energy-momentum tensor operator (EMT) provide a coordinated viewpoint~\cite{Ji:1996nm,Ji:1996ek,Diehl:2003ny,Leader:2013jra,Polyakov:2018zvc} via the gravitational form factors (GFFs) of the EMT.
The lattice-regulated approach to QCD calculations (LQCD) is now able to use the EMT to address both the quark and gluon parton components of the gravitational form factors including the  angular momentum and the mass/energy distribution.  Recent LQCD results include all the form factors of the nucleon including the gluon component of the mass/energy distribution~\cite{Hackett:2023rif} and all the form factors of the pion~\cite{Hackett:2023nkr}.  

There is a close connection between the gravitational form factors of the EMT and the quark and gluon $\langle x \rangle$ moments of  generalized parton distributions (GPDs)~\cite{Ji:1996nm,Radyushkin:1996nd}, and we utilize that for our calculations in this work.   The GPDs are of broader interest~\cite{Diehl:2003ny} because they contain information on imagining of hadrons~\cite{Burkardt:2002hr}, their mass decomposition~\cite{Lorce:2018egm,Hatta:2018sqd}, and cross sections for hard exclusive reactions~\cite{Ji:1996nm,Radyushkin:1997ki} that can be measured at facilities such as Jefferson Lab and an Electron Ion Collider.  The lowest Mellin moments of hadron GPDs at $Q^2=0$ relate to static properties such as the electric charge, magnetic moment, axial charge, and quark spin $S_{\rm q}$.  From the $Q^2$ derivative, it is possible to obtain the quark and gluon parton shares of total angular momentum $J$ of the nucleon~\cite{Tandy:2023zio}. 
Other recent theoretical approaches and models for the parton content of GFFs include an approach from the Instanton Liquid Model~\cite{Liu:2024jno}; the light-front Hamiltonian method with an explicit one-gluon Fock space component~\cite{Xu:2022yxb};  a contact quark-quark interaction model~\cite{Sultan:2024hep}) where the constituent quarks are structureless and the gluon contribution is not accessible; and  an algebraic GPD model~\cite{Raya:2021zrz} in which gluon contributions are only generated by scale evolution.

Here we exploit the close mathematical connection between hadron matrix elements of the EMT and generalized momentum fraction moments to obtain the parton decomposition of the EMT for nucleon and pion  from the Dyson-Schwinger equation approach (DSE) using methods previously applied for standard parton momentum fractions~\cite{Tandy:2023zio}.   
We use the  ladder-rainbow (RL) truncation that selects an infinite subset of gluon emission and absorption processes anticipating that the quark dressing mechanism, so strong because of dynamical chiral symmetry breaking, will generate most of the gluon contributions to parton PDFs and GPDs at the model scale. 
That  same DSE-RL approach has previously proven very successful and efficient for ground state masses, decay constants, and electromagnetic form factors~\cite{Bashir:2012fs,Cloet:2013jya,Tandy:2014hja,Horn:2016rip} especially for light quark pseudoscalar and vector mesons~\cite{Maris:1999nt,Maris:2000sk}. 

\section{The Energy-Momentum Tensor and the Energy Distribution in Hadrons } 
The object of interest is the hadronic matrix element \mbox{$ \langle P^\prime |T^{\mu \nu}(0) | P \rangle$} where \mbox{$ P = K - Q/2$}, \mbox{$ P^\prime = K + Q/2$} with $Q$ spacelike and with hadronic states  covariantly normalized as \mbox{$\langle P^\prime | P \rangle =$} \mbox{$ 2 E \, (2 \pi)^3 \, \delta^3(P^\prime - P) $}.
The QCD Energy-Momentum Tensor (EMT), in the symmetric Belinfante~\cite{Belinfante:1962zz} form, is given by the operator \mbox{$T^{\mu \nu}(x)= $}  \mbox{$ T_q^{\mu \nu}+T_g^{\mu \nu}$}, where
\begin{align}
T_q^{\mu \nu}(x) = \frac{i}{4}\,\bar{q}(x) \, \left[ \LRD^\mu \gamma^\nu + \LRD^\nu \gamma^\mu \right]\, q(x)~,
\label{eq:EMTq(x)}
\end{align}
and 
\begin{align}
T_g^{\mu \nu}(x) = G^{\mu \alpha}(x)\, {G_{\alpha}}^{\nu}(x) + \frac{g^{\mu \nu}}{4}\, G^{\alpha \beta }(x)\, G_{\alpha \beta}(x) ~,
\label{eq:EMTg(x)}
\end{align}
where we have eliminated the \mbox{$ g^{\mu \nu }\, \bar{q}(x) \,\left[ \Dslash - m \right]\, q(x) $} term via the equations of motion. In the above we use a standard notation~\cite{Leader:2013jra} \mbox{$ \LRD^\mu = \RD^\mu -\LD^\mu$} with  \mbox{$ \RD^\mu = \Rpartial^\mu- i g A^\mu $}  and  \mbox{$ \LD^\mu = \Lpartial^\mu + i g A^\mu $}, where the gluon field $A$ includes the flavor factor.
Since the Hamiltonian operator is $\int \! d^3 x \,T^{0 0}(x)$, information on how  mass/energy associated with parton $\rm p $ is distributed in a hadron is expressed by the normalized 3-space distribution 
\begin{align}
{\mathcal E}_{\rm p}(\vec{Q}^2) = \frac{ \left\langle P'  \middle| T_{\rm p }^{0 0}(0) \middle| P  \right\rangle}{2\, E(\vec{Q}^2)}~.
\label{eq:EnFF}
\end{align}
\begin{figure}[tbp]
\centering\includegraphics[width=73mm,height=70mm]{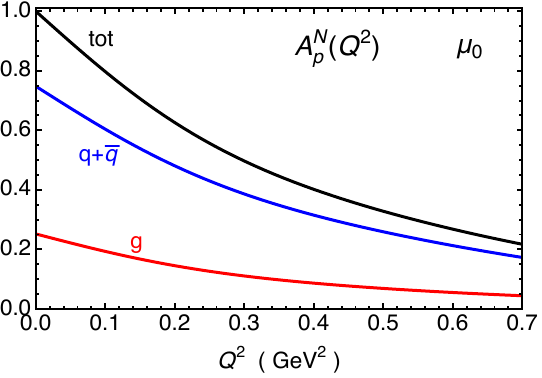}\hspace*{1mm}
\centering\includegraphics[width=73mm,height=70mm]{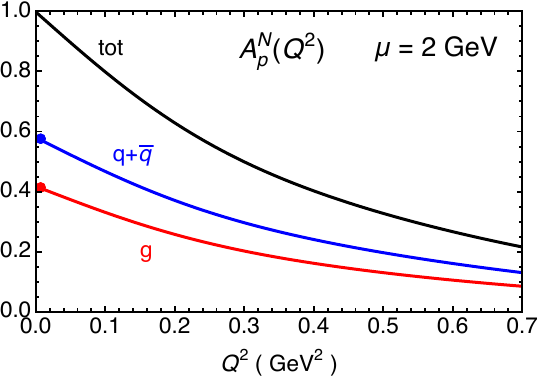}
\caption{{\it Left panel:} The gravitational form factor $A^N(Q^2)$ of the nucleon at the model scale  
showing quark ($q + \bar q$) and gluon contributions. 
{\it Right panel:} The same but at scale \mbox{$\mu = 2$~GeV}. 
The filled circles are the values of $A_{\rm q}(0), \, A_{\rm g}(0)$ that result from a recent global data analysis~\cite{Hou:2019efy}.
}
\label{fig:A_Q2_N_mu0_2GeV}  
\end{figure}

\subsection{The EMT Gravitational Form Factors and GPDs} 
We will utilize two different basis vectors: the light-like longitudinal vector \mbox{$n^\mu = (1;\,\vec{0}_T,\,-1)$} and the rest frame  vector \mbox{$e^\mu = (1;\,\vec{0}_T,\,0)$}.  We take $Q$ to be transverse, \mbox{$ Q \CDOT n = Q \CDOT K = $}  \mbox{$Q \CDOT e $},  where \mbox{$ K =  (P^\prime +P)/2=$} \mbox{$(K^0 ,\vec{0})$}.  The mass-shell condition yields \mbox{$(K^0)^2 = $}  \mbox{$E^2(\vec{Q}^2) = (M^2 + \vec{Q}^2/4 )$}.  For a spin-0 hadron like the pion the  matrix element of  $ T^{\mu \nu}_{\rm q/g}(0) $ has the general form~\cite{Ji:1996nm,Polyakov:2018zvc} 
\begin{align}
\langle \pi(P^\prime) | T^{\mu \nu}_{\rm q/g}(0) | \pi(P) \rangle = A_{q/g}^\pi(Q^2)\, 2 K^\mu \, K^\nu  -  D_{q/g}^\pi(Q^2)\, \frac{Q^2\,\mathcal{T}^{\mu \nu}(Q) }{2}  + \bar{C}_{\rm q/g}^\pi(Q^2)\, 2\, M^2 \, g^{\mu \nu}~,
\label{eq:EMT_Pi_genF}
\end{align}
with the transverse projector being  \mbox{$ \mathcal{T}^{\mu \nu}(Q) = $} \mbox{$ g^{\mu \nu}  - Q^\mu Q^\nu /Q^2 $}.  The  $A_{q/g}^\pi(Q^2)$ are the parton components of the prominent gravitational form factor whose limit \mbox{$A_{q/g}^\pi(0) = \langle x \rangle_{q/g}$} yields the parton momentum fractions associated with the DIS process.  The other gravitational form factors relate to the distribution of internal pressure and stress.
The nucleon version of \Eq{eq:EMT_Pi_genF}, with spinors normalized as \mbox{$\bar{u}(P)  u (P) = 1$}, and thus \mbox{$\bar{u}(P^\prime)  u(P) = $}  \mbox{$E(\vec{Q}^2)/M$},  is~\cite{Ji:1996nm,Polyakov:2018zvc} 
\begin{align}
\langle N(P^\prime) | T^{\mu \nu}_{\rm q/g}(0) | N(P) \rangle = \bar u(P^\prime) \big\{ A^N_{q/g}(Q^2) \, 2\, K^\mu K^\nu + J^N_{q/g}(Q^2) \,t^{\mu \nu}(K,Q,\sigma)  \nonumber  \\ 
- \, D^N_{q/g}(Q^2) \,  \frac{Q^2\,\mathcal{T}^{\mu \nu}(Q) }{2}  + \bar{C}_{\rm q/g}^N(Q^2) \, 2\, M^2\, g^{\mu \nu} \big\} u(P) ~.
\label{eq:EMT_N_genF}
\end{align}
In the nucleon  case the state labels $P^\prime , P$ are to be understood to include spin projections $s^\prime , s$. 
Here the tensor $ t^{\mu \nu}$ is \mbox{$ i \,K^{ \{ \mu }  \sigma^{\nu \} \alpha} Q_\alpha = $} \mbox{$ i\, (K^\mu \sigma^{\nu \alpha} +K^\nu \sigma^{\mu \alpha})\, Q_\alpha $},
and  \mbox{$J_{\rm p}^N(Q^2) \equiv $} \mbox{$[A^N_{\rm p} + B^N_{\rm p} ]/2$} describes the distribution of total angular momentum contributed by  parton $p$.  By symmetry \mbox{$\partial_\mu T^{\mu \nu}(x) = 0$} and thus \mbox{$\Sigma_p \bar{C}_{p}(Q^2) = $}  \mbox{$0$}.    Also by symmetry, the  \mbox{$ Q^2 \to 0 $} limit \mbox{$ \langle P | T_{\rm q + g}^{+ +}(0) | P \rangle/ 2\, M^2=$} \mbox{$ 1 $} yields the sum rule \mbox{$ \langle x \rangle_q +  \langle x \rangle_g = 1 $} for nucleon or pion.   A byproduct is that for the nucleon \mbox{$\Sigma_p B^N_p(0) = 0$}. 
The form factors $A_p(Q^2), B_p(Q^2)$  are the parton light-cone  momentum fraction moments (i.e., second Mellin moments) of the GPDs $H_p(x,\xi,Q^2), E_p(x,\xi,Q^2)$ respectively~\cite{Ji:1996nm,Polyakov:2018zvc}  where \mbox{$\xi = $} \mbox{$ - Q \CDOT n/(2K \CDOT n) \to 0$} because here $Q$ is transverse.   

The various form factors can be isolated if one has the contraction of the EMT with sufficient independent kinematic tensors.  Sufficient are the three symmetric tensors  $n_\mu n_\nu$  (giving $T^{+ +}_{\rm p}$)  and $e_\mu e_\nu$ (giving $T^{0 0}_{\rm p}$), and \mbox{$ Q_\mu Q_\nu/Q^2$}.  Here we report results from use of the first two. For a spin-$\frac{1}{2}$ hadron the general form of $\langle T^{+ +}_{\rm p} \rangle$ is
\begin{align}
n_\mu \langle N(P^\prime) | T^{\mu \nu}_{\rm p}(0) | N(P) \rangle n_\nu  =  2 M \,K^0(Q) \;\bar{u}_{s'}(P')  \left[  \nslash\, A^N_{\rm p}(Q^2)
  +   \frac{i\sigma^{n \alpha}Q_\alpha}{2M}\,  B^N_{\rm p}(Q^2)   \right] u_{s}( P ) ,
\label{eq:gen_form_JQFT}
\end{align}
which allows the separate identification of $A^N_{\rm p}$ and $B^N_{\rm p}$ upon using the $s^\prime s$ dependence.   For the pion, the general form of $\langle T^{+ +}_{\rm p} \rangle$ is simply \mbox{$ n_\mu \langle \pi(P^\prime) | T^{\mu \nu}_{\rm p}(0) | \pi(P) \rangle n_\nu  = $} \mbox{$ 2 (K^0)^2 \,A^{\pi}_{\rm p}(Q^2)$}.  The contraction with $e_\mu e_\nu$  generally involves a linear combination of all form factors.  A further contraction with $Q_\mu Q_\nu/Q^2$ will isolate $ \bar{C}_{\rm p}(Q^2) $ and allow a complete extraction.   We don't consider contraction with $Q_\mu Q_\nu/Q^2$ here; but at \mbox{$Q^2 = 0$} we can extract  the $\bar{C}_{\rm p}(0)$ because $D_{\rm p}(0)$ does not contribute.

\section{Towards  calculation of  $ a_\mu T^{\mu \nu} a_\nu $}
With  $T^{a a}$ denoting \mbox{$a_\mu T^{\mu \nu} a_\nu$}, where vector $a^\mu$ can be the above $n^\mu$ or $e^\mu$, the QCD matrix element of $T^{a a}_{\rm q}(0) $ from \Eq{eq:EMTq(x)} can be rearranged by translating the gluon field part of the covariant derivative $\LRD^a $ into a derivative of a Wilson line integral, resulting in 
\begin{align}
\left\langle P'  \middle| \,\bar{q}(0) \, \frac{i}{2} \, \LRD^a \, \aslash \, q(0)  \middle| P  \right\rangle = \int d^4z\, \int_k\,  e^{i k\cdot z} \, i \partial^a(z) \left\langle P'  \middle| \bar{q}(-\frac{z}{2}) W (-\frac{z}{2}, \frac{z}{2})\, \aslash q(\frac{z}{2}) \middle| P  \right\rangle~,
\label{eq:C}
\end{align}
where \mbox{$\int_k = $} \mbox{$ \int d^4k/(2 \pi)^4$}.
Here the Wilson line integral \mbox{$ W (a, b) = P e^{i g \int^a_b dy \cdot A(y)} $} restores explicit  gauge invariance to this explicitly non-local form.   Use of  integration by parts to reverse the action of  $i \partial^a(z)$ produces a factor $k \CDOT a$ leading to the matrix element of \Eq{eq:EMTq(x)} expressed in the  form 
\begin{align}
\frac{\left\langle P'  \middle| T_{\rm q}^{a a}(0) \middle| P  \right\rangle}{2 (K^0)^2} =  \frac{1}{2K^0} \! \int_k\,(\frac{k \CDOT a}{K^0}) \int \!d^4z \,e^{i k\cdot z} \, \left\langle P'  \middle| \bar{q}(-\frac{z}{2}) W (-\frac{z}{2}, \frac{z}{2})\, \aslash q(\frac{z}{2}) \middle| P  \right\rangle~.
\label{eq:D}
\end{align}
With the choice  \mbox{$a^\mu \to n^\mu$} this yields $\langle x \rangle_{\rm q}$ associated with the DIS process as well as its $Q^2 > 0$ generalization.   
The corresponding gluon parton matrix element of \Eq{eq:EMTg(x)} can be expressed in the similar form
\begin{align}
\frac{\left\langle P'  \middle| T_{\rm g}^{a a}(0) \middle| P  \right\rangle}{2 (K^0)^2} =  \frac{1}{2K^0} \! \int_k  (\frac{k \CDOT a}{K^0}) \!\int \!d^4z \,\frac{e^{i k\cdot z}}{k \CDOT a}   \left\langle P'  \middle|  G^{a \nu}(-\frac{z}{2}) \, {G_{\nu}}^a(\frac{z}{2}) + \frac{a\CDOT a}{4}\, G^{\mu \nu }(-\frac{z}{2})\, G_{\mu \nu}(\frac{z}{2})    \middle| P  \right\rangle,
\label{eq:G}
\end{align}
where \mbox{$G^{a\, \nu}(z) = a_\mu \,G^{\mu\, \nu}(z)$}. The Wilson line integral is unity for this lowest physical gluon moment due to cancellation of $k\CDOT a$.
In these expressions the $\int \! d^4z $ result defines the representation of the process in terms of parton momentum $k$; then the $\int \! d^4k$ implements the momentum fraction $\langle k\CDOT a/K\CDOT a \rangle$  expectation value.  These are generalized momentum fractions related to the component in direction $a$; only  the special case $a\! \to\! n$ yields the lightcone momentum fraction associated with a DIS process. In that case $g^{n n}\! \to \!0 $ and both the gluon and quark expressions in \Eq{eq:D} and \Eq{eq:G}  yield the standard QCD light-cone momentum fraction expressions~\cite{Diehl:2003ny} accessed in the DIS process.  We draw this analogy of  $\left\langle P'  \middle| T_{\rm p}^{a a}(0) \middle| P  \right\rangle$ with DIS momentum fractions to enable the application of such  established calculation methods to produce EMT results.  

\begin{figure}[tbp]
\centering\includegraphics[width=73mm,height=70mm]{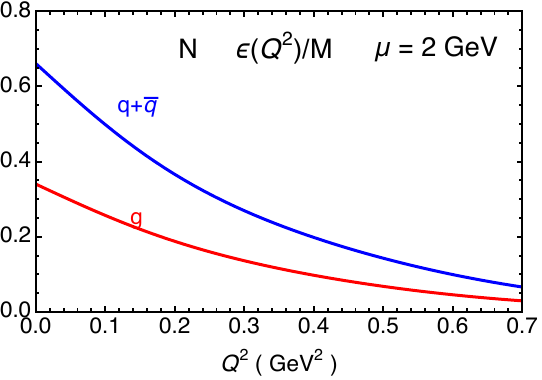}\vspace*{1mm}
\centering\includegraphics[width=73mm,height=70mm]{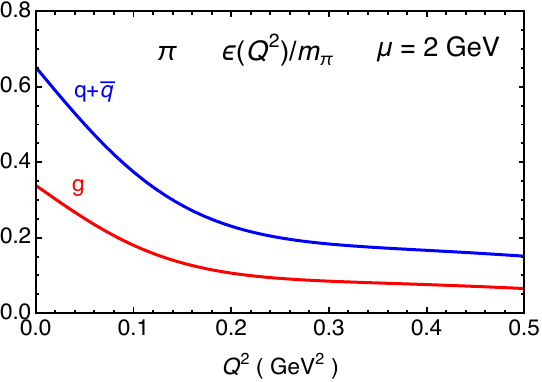}
\caption{{\it Left panel:} The unit-normalized energy distribution ${\mathcal E}(Q)/M$ of the nucleon at scale \mbox{$\mu = 2~{\rm GeV}$}, showing the quark and gluon contributions.  
{\it Right panel:} The pion unit-normalized energy distribution.   
}
\label{fig:EnDen_N_Pi_2GeV}  
\end{figure}

\section{Dynamical Approach} 
The previous definitions employed Minkowski metric; from here on we adopt Euclidean metric in order to apply the Rainbow-Ladder truncation of the DSE approach to non-perturbative aspects of hadron physics.\footnote{In Euclidean metric we employ $a_4 = i a^0$ for any space-time vector, including $n$, while  \mbox{$\{\gamma_\alpha, \, \gamma_\beta\} =2 \delta_{\alpha \beta} $}  with $\gamma_4 = \gamma^0$.   Hence \mbox{$\aslash \to -i \aslash $} while \mbox{$a \CDOT b \to - a \CDOT b $}.  }  This has successfully described many hadron properties~\cite{Bashir:2012fs,Cloet:2013jya,Tandy:2014hja,Horn:2016rip}, in particular it was previously used to explore dynamical relations between quark and gluon contributions to PDFs angular momentum of the proton~\cite{Freese:2021zne,Tandy:2023zio}.  

The Rainbow-Ladder  truncation of the QCD matrix elements  in \Eq{eq:D} and \Eq{eq:G} yield for parton type ${\rm p}$~\cite{Freese:2021zne,Tandy:2023zio}
\begin{align}
\langle \frac{k\CDOT a}{K\CDOT a} \rangle_{\rm p}(Q^2) = {\rm tr}\! \int_p \! S(p^f)\, \Gamma^{a}_{\rm p}(p,Q)\, S(p^i) \,{\mathcal M}(p,Q;\!P), 
\label{eq:DSE_m1_N_Taa}
\end{align}
where  \mbox{$p^{f/i} = p\, \pm Q/2$} and the trace is over Dirac and color indices.
For the pion  ${\mathcal M}(p,Q;\!P)$ is $\Gamma_\pi(p^i \!-P/2; P) \, S(p^i \! -P) \, \bar{\Gamma}_\pi(p^f \!-P^\prime/2; -P^\prime)$.   In the nucleon case, ${\mathcal M} $ contains a dependence on initial and final spin projections and the flavor of the active parton; the employed model is described later.   In \Eq{eq:DSE_m1_N_Taa}  the $ \Gamma^{a}_{\rm p}(p,Q)$ are DSE-RL dressed vertices that carry the  second Mellin moment  $\langle k\CDOT a/K\CDOT a \rangle$ information generated by the quark-in-quark aspect ($ \Gamma^{a}_{\rm q}$)  or the gluon-in-quark aspect ($ \Gamma^{a}_{\rm g}$).   The strength of dynamical chiral symmetry breaking for light quarks suggests the quark dressing mechanism produces most of the strength of the gluon parton behavior; the  obtained results support  this.

For realistic numerical work, the dressed quark propagator $S(p)$ is obtained from the quark Dyson-Schwinger equation of QCD in Rainbow-Ladder truncation, which is 
\begin{align}
S^{-1}(p) = Z_2\,S_0^{-1}\!(p) - 
\!\int_k \frac{\lambda^a}{2} \gamma_\mu \,\mathcal{K}_{\mu \nu}(q)\,  S(k)\, \frac{\lambda^a}{2} \gamma_\nu,
\label{SigmaRL}
\end{align}
where \mbox{$ S_0^{-1}(p) = i \pslash + Z_m m_r $}.  The general form of the solution is 
$S^{-1}(p)  = i \pslash \,A(p^2, \mu^2) + B (p^2, \mu^2)$, and the renormalization constants ($Z_2, Z_m$) produce \mbox{$A \to 1$} and \mbox{$B \to m_r$} at the renormalization scale $\mu$.  The  standard Landau gauge DSE-RL interaction kernel~\cite{Maris:1999nt,Nguyen:2011jy,Qin:2011dd} that generates quark propagators, BSE vertices and meson bound states is \mbox{$ \mathcal{K}_{\mu \nu}(q) = {\mathcal K}(q^2) \,\mathcal{T}_{\mu \nu} (q)$} where $\mathcal{T}_{\mu \nu}(q) $ is the transverse projector and ${\mathcal K}(q^2)$ is the model's non-perturbative generalization of the bare kernel \mbox{${\mathcal K}^0(q^2) = g^2/q^2$}.  This DSE-RL kernel correlates a large amount of hadron physics~\cite{Bashir:2012fs,Cloet:2013jya,Tandy:2014hja,Horn:2016rip} and is given by   
\begin{align}
{\mathcal K}(q^2) = D_{\rm RL}^2  \, {\rm e}^{-q^2/\omega^2} 
+ {\mathcal F}(q^2)\, 4\pi\, \tilde{\alpha}_s(q^2)  \, .
\label{eq:RL_Kernel}
\end{align}
Here $\tilde{\alpha}_s(q^2)$ denotes a continuation of the 1-loop $\alpha_s(q^2)$ to provide smooth non-singular coverage for the entire domain of $q^2$.   The first term of \Eq{eq:RL_Kernel} implements the infrared enhancement due to dressing effects, while the second term, with \mbox{${\cal F}(q^2)=$} \mbox{$(1 - \exp( -q^2/(1~{\rm GeV^2}) ) )/q^2 $}, connects smoothly with the 1-loop renormalization group behavior of QCD.  

The defining Bethe-Salpeter equation for the dressed vertices $ \Gamma^{a}_{\rm p}$  is~\cite{Nguyen:2011jy,Bednar:2018mtf,Freese:2021zne}
\begin{align}
\Gamma^{a}_{\rm p}(p,Q) =  \Gamma^{a}_{\rm p,D} (p,Q) 
- \int_\ell \frac{\lambda^a}{2} \gamma_\mu \mathcal{K}_{\mu \nu}(q)\, S(\ell^f)\, \Gamma^{a}_{\rm p}(\ell,Q)\, S(\ell^i)\, \frac{\lambda^a}{2} \gamma_\nu~, 
\label{eq:Gamma_BSE}
\end{align}
where \mbox{$ \ell^{f/i} = \ell \pm Q/2$}, \mbox{$q = p - \ell $}, and  $\mathcal{K}_{\mu \nu}(q)$ is the DSE-RL kernel.  
The inhomogeneous vertex terms are those in ~\Refs{Freese:2021zne,Tandy:2023zio} but here generalized to accommodate the more general basis vector $a$.  In the quark parton case  \mbox{$\Gamma^{a}_{\rm q,D} (p,Q) =  $}  \mbox{$Z_2\,(-i \aslash)\,(\frac{ p \CDOT a}{K^0}$}), while the gluon parton case yields
\begin{align}
\Gamma^{a}_{\rm g,D} (p,Q) \!= \!\frac{4}{3} \!\int_q  \left(\!\frac{q \CDOT a}{K^0}\! \right)  \gamma_\mu \, \Delta_{\rm g}(q^f) \, \hat{D}_{\mu \nu}(q,Q) \, \Delta_{\rm g}(q^i) \,S(p\!-\!q) \, \gamma_\nu ,
\label{eq:Inhom_GVertm}
\end{align} 
where $S(p\!-\!q)$ is the quark propagator.    The inhomogeneous gluon-in-quark vertex involves $\Delta_{\rm g}(q)$ which is determined by the corresponding $Q=0$ kernel for the momentum fraction calculation.   In particular  \mbox{$\Delta_{\rm g}(q)^2 = $} \mbox{$ {\mathcal K}_{\rm g}(q^2) $} is the  non-perturbative generalization of $g^2/q^4$ given  in ~\Refs{Freese:2021zne,Tandy:2023zio}.  \Eq{eq:Inhom_GVertm} involves the tensor is \mbox{$ \hat{D}_{\mu \nu} =  $}  \mbox{$\mathcal{T}_{\mu \alpha}(q^f) \, \tilde{Y}_{\alpha \beta}(q^f, q^i) \, \mathcal{T}_{\beta \nu}(q^i)  $} with $\mathcal{T}_{\mu \alpha}$ being the transverse projector, as well as  the tensor $ \tilde{Y}_{\alpha \beta}$  which is
\begin{align}
\tilde{Y}_{\alpha \beta}   = \frac{2}{q \CDOT a} \Big(  q^f \!\CDOT a \, q^i \!\CDOT a\, \delta_{\alpha \beta}   -  q^f \!\CDOT a \, q^i_\alpha  \, a_\beta  
+  {q^f} \!\CDOT {q^i} \, a_\alpha \, a_\beta  -  a_\alpha \, q^f_\beta \, q^i  \!\CDOT a \Big)  + \frac{a\CDOT a}{q\CDOT a} \Big( q^i_\alpha  \, q^f_\beta 
- {q^f} \!\CDOT {q^i} \, \delta_{\alpha \beta} \Big)~,
\label{eq:Inhom_Ytensor}
\end{align}
where  \mbox{$q^{f/i} = q\, \pm Q/2$}.  The numerator terms are bilinear in momentum and this structure derives from \Eq{eq:G} being bilinear in the gluon field tensor and then subsequent truncation to ladder-rainbow structure as needed here.  The second term here originates from the term of $T^{\mu \nu}$ proportional to $g^{\mu \nu} $, that is the second term of \Eq{eq:G}.  

\begin{figure}[tbp]
\centering\includegraphics[width=73mm,height=70mm]{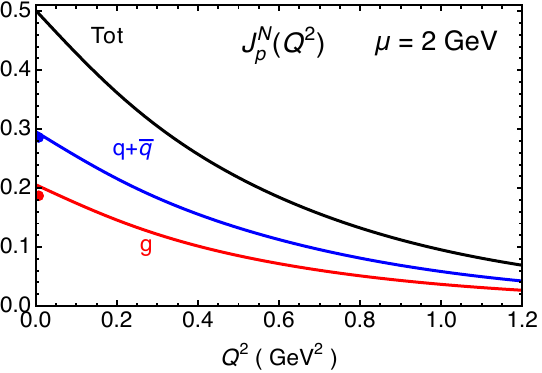}\hspace*{1mm}
\centering\includegraphics[width=73mm,height=70mm]{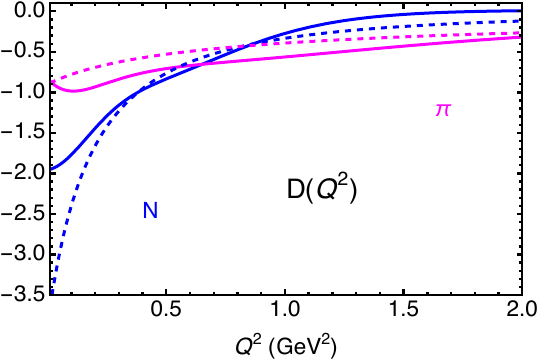}
\caption{  {\it Left panel:}  The angular momentum distribution $J(Q)$ of the nucleon at scale $2$~GeV, showing quark ($q+\bar{q}$) and gluon contributions.    The filled circles are the values of $J_{\rm q}(0), \, J_{\rm g}(0)$ from a recent LQCD calculation~\cite{Alexandrou:2020sml}.
{\it Right panel:} The gravitational form factor $D(Q)$ of the nucleon and pion summed over partons.  The dashed curve for the nucleon is the dipole fit obtained in the recent LQCD calculation~\cite{Hackett:2023rif}, and the dashed curve for the pion is the monopole fit obtained in an earlier LQCD caculation~\cite{Hackett:2023nkr}.
}
\label{fig:DFF_N_Pi_2GeV}  
\end{figure}

The DSE-RL gluon-in-quark  kernel ${\mathcal K}_{\rm g}(q^2)$ generates the gluon momentum fraction and its strength is set~\cite{Tandy:2023zio} so that the pion's  model scale momentum fractions $\langle x \rangle_{\rm q} $ and $ \langle x \rangle_{\rm g} $ lead  to a minimized RMS departure from global data analysis results at upper scales.  Due to the Landau gauge of the DSE-RL approach and its lack of an explicit Wilson line integral, we employ the 1-loop estimated compensation~\cite{Tandy:2023zio} of a 7\% decrease of  $\langle x \rangle_{\rm q}(\mu_0)$ and a 27\% increase of $\langle x \rangle_{\rm g}(\mu_0)$ to maintain the fit to data at higher scales.  Kernel parameters are summarized in Ref.~\cite{Tandy:2023zio}.   
After the pion is used to fix the gluon-in-quark kernel parameters, that kernel is applied unchanged to the nucleon.  
The deduced model scales are \mbox{$\mu^\pi_0=$}  \mbox{$0.9~{\rm GeV}$}, \mbox{$\mu^N_0=$}  \mbox{$0.64~{\rm GeV}$}. 

The nucleon matrix element in \Eq{eq:DSE_m1_N_Taa} involves a \mbox{$Q \geq 0$} generalization of the model used previously to study $\langle x \rangle$  sharing between quarks and gluons~\cite{Freese:2021zne}.  Here
\begin{align}
\mathcal{M}^f_{s^\prime, s}(p,Q;\!P)  =N \, \left[ u_{s}(p^i) \, \bar{u}_{s^\prime}(p^f) \right] \rho_{f,s^\prime, s} \,\tilde{f}(q^\prime)\, \tau(P\!- \!p^i)\, \tilde{f}(q)~,
\end{align}
where \mbox{$p^{f/i} = p \pm Q/2$}  and $\rho_{f,s^\prime, s}$ is the relevant spin-flip quark density for  flavor $f$ in the pure SU(6) spin-isospin proton state.  The diagonal spin probabilities yield the quark numbers  \mbox{$ \small\sum_s \, \rho_{f,s,s} = n_f $}, while  the off-diagonal probabilities are \mbox{$\rho_{u,\uparrow,\downarrow} =\frac{4}{3} $} and \mbox{$\rho_{d,\uparrow,\downarrow} = - \frac{1}{3} $}.  
Here \mbox{$q^\prime = p^f \!- \!P^\prime \!/3$} and \mbox{$q = p^i \!- \!P/3$} are the relative momenta of the active quark and  spectator pair in the final and initial state respectively, while  \mbox{$\tilde{f}(k) = 1/( k^2 + R^2)^3 $} describes the momentum dependence while reproducing the high-$x$ exponent behavior of the produced valence PDF $q(x)$.  The function $\tau(s)$ 
represents the di-quark propagator $1/( s^2 + M_{\rm D}^2) $ with proper time regularization of the deep infrared domain. The normalization $N $ is fixed by the valence quark numbers $n_f$.   The parameters were fit to results from Faddeev calculations of the nucleon~\cite{Bednar:2018htv}.

\section{Results} 
In \Fig{fig:A_Q2_N_mu0_2GeV} we display the obtained nucleon distributions $A_{\rm q/g}(Q^2)$ at the model scale \mbox{$\mu^N_0=$}  \mbox{$0.64~{\rm GeV}$} and at $2$~GeV.
The gluon contribution at the model scale is not insignificant; it is  larger than half of the value at 2~GeV.  Furthermore these fractions at $Q^2 =0$ are the parton lightcone momentum fractions associated with the DIS process.   The filled circles are the values of $A_{\rm q}(0), \, A_{\rm g}(0)$ that result from a recent global data analysis~\cite{Hou:2019efy}.

In \Fig{fig:EnDen_N_Pi_2GeV} we display the unit-normalized mass/energy distributions ${\mathcal E}_{\rm p}(Q^2)/M$ for parton $p$ of the nucleon and pion.  A consequence of the definition of ${\mathcal E}_{\rm p}(Q^2)/M$ from \Eq{eq:EnFF}, and its relation to the other form factors derivable from \Eq{eq:EMT_Pi_genF} and \Eq{eq:EMT_N_genF}, is that the parton fractions  ${\mathcal E}_{\rm p}(0)/M$ are not identical to  $A_{\rm p}(0) $.  The difference is due to the  non-conservation form factors $\bar{C}_{\rm q/g}^\pi(0) $ and for \mbox{$\mu = 2~{\rm GeV}$ } we estimate
\begin{align}
\begin{tabular}{cc||cc} 
         ${\bar C}^N_{\rm q}(0)$  & ${\bar C}^N_{\rm g}(0)$  &    ${\bar C}^\pi_{\rm q}(0)$  & ${\bar C}^\pi_{\rm g}(0)$    \\
\hline
         -0.08   &   0.08   &    -0.02  &  0.02      \\
\end{tabular}~.
\label{eq:Cbar0_N_pi}
\end{align}

In \Fig{fig:DFF_N_Pi_2GeV} we display the obtained angular momentum form factor $J_{\rm p}(Q^2)$ for parton $p$ of the nucleon. Since $J = (A + B)/2$ and $A$ is strongly dominant, the parton fractions for $J$ are almost the same as those for $A$ shown in \Fig{fig:A_Q2_N_mu0_2GeV}.   Also shown is the total form factor  $D(Q^2)$.  This is obtained by using summed parton contributions to $\langle T^{0 0} \rangle$,  $A$,  $J$, and taking the appropriate linear combination.    The result should be scale invariant and this approach repeated at $\mu_0$ and $2$~GeV produces nearly identical results. 

We identify radii associated with each parton distribution $f(Q^2) $ in the conventional way: \mbox{$ R^2 = $} \mbox{$- 6  f^\prime(0)/f(0)$}. The radii obtained for the distributions associated with $A_{\rm p}(Q^2)$ and ${\mathcal E}_{\rm p}(Q^2)$ at $2$~GeV are displayed in \Table{tab:Radii_2GeV}.   The radii of the gluon contribution to the mass/energy distributions are seen to be essentially equal to the net quark radii for both nucleon and pion.  Another major pattern that emerges is that the distribution radii associated with form factor $A_{\rm p}(Q^2)$ are uniformly less than those of $ {\mathcal E}_{\rm p}(Q^2)$, especially for the pion.  This is mainly due to the denominator arising from the covariant normalization of states.  From \Eq{eq:EnFF} the radii  associated with ${\mathcal E}_{\rm p}$ and $ T_{\rm p }^{0 0}$ are related by \mbox{$R^2_{\mathcal E_{\rm p}}  = $}  \mbox{$R^2_{\rm T_{\rm p}} + 3/4 M^2$}; for a light hadron mass $M \to m_\pi$ this represents a significant increase.  The distribution associated with $T_{\rm p}^{0 0}(Q^2)$ is influenced by several form factors besides $A_{\rm p}(Q^2)$, but in general the radii have the hierachy $R_{\rm A} < R_{\rm T} < R_{\mathcal E}$; the form factor $A_{\rm p}(Q^2)$ does not well represent the distribution of mass/energy.

In \Table{tab:Radii_2GeV} the obtained gluon mass/energy distribution radius $ R_{ {\mathcal E}_{\rm g}}$ for the nucleon compares well to the recent LQCD result ~\cite{Hackett:2023rif} $0.81$~fm,  and the recent analysis~\cite{Meziani:2024cke} of data yielding $0.778$~fm.   Both of these works ignored the influence of $ {\bar C}_{\rm g}(Q^2)$ in their analysis.   Here its effects have been included because we  obtain ${\mathcal E}_{\rm g}$ directly from the total QCD matrix element $ T^{0 0}_{\rm g}$.  The obtained ratio $ R_{\rm g}/ R_{\rm q} = 1.22$ for the pion form factor $A$ compares well with the value $1.1$ obtained recently from LQCD~\cite{Hackett:2023nkr}, as well as the value $1.31$ obtained via an algebraic GPD model~\cite{Raya:2021zrz}.  Also shown in \Table{tab:Radii_2GeV} are the quark radii from both $A_{\rm q}$ and ${\mathcal E}_{\rm q}$ expressed as a ratio to the empirical charge radii. 

After obtaining radii for the total (summed parton) distributions ${\mathcal E}(Q^2)$ and $A(Q^2)$ the value of parton total $D(Q^2 \!= \!0)$ can be obtained from those total radii. One finds \mbox{$R_{ {\mathcal E}_\pi}^2 \!= $} \mbox{$ R_{{\rm A}_\pi}^2 \!- \!(3/2M^2)(1/2 + D_\pi(0) ) $} and \mbox{$R_{ {\mathcal E}_{\rm N}}^2 \!= $} \mbox{$ R_{{\rm A}_{\rm N}}^2 \!- \!(3/2M^2)D_{\rm N}(0)  $}.  This yields \mbox{$D_\pi(0) = -0.89$} and  \mbox{$D_{\rm N}(0) = -1.73$} at the scale \mbox{$\mu = 2$}~GeV; these values provide a consistency check with those shown in \Fig{fig:DFF_N_Pi_2GeV}.  

\begin{table}[tbp]
%
\begin{tabular}{c|c||c|}\hline
 \hspace*{23.5mm} &  \hspace*{26.0mm} $ A$ \hspace*{26.0mm} &  \hspace*{25.4mm} $ {\mathcal E} $ \hspace*{25.4mm}  \\
\hline 
\end{tabular}\\
\begin{tabular}{c| cc|c|c||cc|c|c|} \hline
    & \hspace{0mm} $ R_{\rm q}$ \hspace{1mm} &  $ R_{\rm g} $ \hspace{1mm} & \hspace{1mm}   $ R_{\rm g}/R_{\rm q} $ \hspace{0mm}  &  $ R_{\rm q}/R_{\rm ch} ^{\rm expt}$   & $ R_{\rm q}$ \hspace{0mm} &  $ R_{\rm g} $ \hspace{0mm} & \hspace{0mm}   $ R_{\rm g}/R_{\rm q} $  &  $ R_{\rm q}/R_{\rm ch} ^{\rm expt}$  \hspace{0mm}  \\
\hline
\rule{0em}{3ex}                                         
    $\pi(\mu = 2~{\rm GeV})$     &  0.397  &  0.485    &   1.22    &  0.60   &  1.178     & 1.197     &   1.017  &   1.79  \\
 \rule{0em}{3ex} 
     $N(\mu = 2~{\rm GeV})$     &   0.627  &   0.655    &   1.044   &  0.75   &  0.766   &   0.765   &   0.998  &  0.912   \\
\hline
\end{tabular}
\caption{ Radii (in fm) obtained  
for the separate quark and gluon distributions $ A_{\rm p}(Q), \, {\mathcal E}_{\rm p}(Q) $ at scale $\mu = 2~{\rm GeV}$.   Ratios use empirical charge radii \mbox{$R^\pi_{\rm ch},\, R^N_{\rm ch} = 0.659, 0.843$}~fm.
}
\label{tab:Radii_2GeV}
\end{table}

\section{Summary}
We have employed the model based on the Dyson-Schwinger Equations in Rainbow-Ladder truncation to  assess  whether the dressing of quarks can, by itself, produce realistic gluon and quark contributions to light-cone momentum fractions, gravitational form factors, mass/energy distributions for the nucleon and pion.     Matrix elements of several kinematic projections of the Energy-Momentum Tensor are directly calculated by utilizing the similarity of its structure  to the generalized  momentum fraction moment of GPDs associated with deep inelastic scattering.  In this exploratory work we have shown this approach to be applicable to a variety of gravitational form factors including the D-term.

The form factors and radii obtained here continue to indicate that the quark dressing mechanism is a strong basis  for the gluon parton contributions. 
An extension of this DSE-RL approach to also calculate the QCD matrix element for $Q_\mu T^{\mu \nu}_{\rm p}  Q_\nu /Q^2$ would yield the $\bar{C}_{\rm p}(Q^2)$ and might be of future interest, as would incorporation of a more realistic treatment of the nucleon state.

%

\end{document}